\documentclass[fleqn,10pt]{wlscirep}
\usepackage[utf8]{inputenc}
\usepackage[T1]{fontenc}

\usepackage{subfiles}

\usepackage{soul, xcolor}
\usepackage{braket}
\usepackage{float}

\title{Autonomous Recognition of Erroneous Raw Key Bit Bias in Quantum Key Distribution}

\author[1,*]{Matt Young}
\author[2]{Marco Lucamarini}
\author[1]{Stefano Pirandola}
\affil[1]{Department of Computer Science, University of York, York, YO10 5DD, United Kingdom}
\affil[2]{School of Physics, Engineering, and Technology, and York Centre for Quantum Technologies, University of York, York, YO10 5DD, United Kingdom}

\affil[*]{matt.young@york.ac.uk}

\begin{abstract}
As Quantum Key Distribution technologies mature, it is pertinent to consider these systems in contexts beyond lab settings, and how these systems may have to operate autonomously.
To begin, an abstract definition of a type of error that can occur with regard to the ratio of bit values in the raw key is presented, and how this has an impact on the security and key rate of QKD protocols.
A mechanism by which errors of this type can be autonomously recognised is given, along with simulated results.
A two part countermeasure that can be put in place to mitigate against errors of this type is also given.
Finally some motivating examples where this type of error could appear in practice are presented to add context, and to illustrate the importance of this work to the development of Quantum Key Distribution technologies.
\end{abstract}
\begin{document}

\flushbottom
\maketitle
\thispagestyle{empty}

\section*{Introduction}

With the advancement of research into Quantum Key Distribution (QKD)\cite{gisin2002quantum, pirandola2020advances}, we are moving ever closer to a world in which QKD systems are a part of real-world infrastructure \cite{shimizu2014performance, amies-king2023feasibility}. 
In such a setting, determining if a QKD system is operating as it should is a task that will need to be performed by a QKD system autonomously. 
In addition to this, downtime of a QKD system should be kept to a minimum, otherwise the operation of critical organisations or infrastructure making use of QKD systems could be severely compromised.

In this work a new scheme is presented to autonomously recognise an error where Bob is no longer able to detect one of the encoding states used by Alice, which manifests as a change in the ratio of bit values in Bob's raw key. 
For brevity, in this work this will be referred to as a state detection error, or SD error.
SD error recognition is done by performing an additional step using the generated raw key, after which Alice and Bob can automatically put a countermeasure in place allowing the QKD system to continue operation, preventing downtime, and autonomously adding redundancy to QKD systems.

To begin, an introduction of the 3-state BB84 protocol, and side-channel attacks are given, which are some of the pre-requisite core concepts used in the main body of this work.
The consequences of failing to recognise a SD error are discussed, before then going on to present the procedure for recognising an erroneous ratio of raw key bits, after which a countermeasure that can be taken by Alice and Bob is given.
Following this, motivating examples of how a SD error could occur are given in the form of a fault of a measurement device, and of a possible side-channel attack.

%\subfile{pa.tex}
\subsection*{3-State BB84}

The original QKD protocol proposed by Bennett and Brassard \cite{bennett2014quantum} (BB84) makes use of 4 quantum states to encode information, $\ket{0}$, $\ket{1}$, $\ket{+}$, and $\ket{-}$, where 
% \begin{equation}
%     \label{eq:phase_basis_states}
%     \ket{\pm}=\frac{1}{\sqrt{2}}(\ket{0}\pm\ket{1})
% \end{equation}
    $\ket{\pm}=\frac{1}{\sqrt{2}}(\ket{0}\pm\ket{1})$.

There exists a specification and security proof of BB84 that only makes use of three of these states \cite{fung2006security}.
For illustrative purposes, the three states $\ket{0}$, $\ket{1}$, and $\ket{+}$ will be considered, but any three of the four BB84 states mentioned above can be used.
There are only two changes from BB84:
\begin{enumerate}
    \item When Alice randomly chooses to prepare a state from the phase-basis (also known as the x-basis), she always prepares the $\ket{+}$ state.
    \item Only instances from the computational-basis (z-basis) are used for key-generation, whilst all instances from the x-basis, and a small fraction from the z-basis are used for parameter estimation.
\end{enumerate}
It has since been shown by other security proofs that the achievable Secret-Key-Rate (SKR) for the 3-State BB84 protocol is almost identical to that of the BB84 protocol, except for situations with really high attenuation \cite{tamaki2014loss, rusca2018security}.
Tamaki et al. \cite{tamaki2014loss} even go on to state that as the two SKRs are so similar, it implies that one of the states in BB84 is redundant.
It is this redundancy that this work makes use of to remove the dependency on the encoding state that is no longer being detected by Bob.
\subsection*{Side-Channel Attacks}

When analysing the security of cyber-security related systems, whether they be classical or quantum, the primary consideration is one of if the underpinning theoretical principles are secure.
However, other avenues for attack open up when considering imperfections not in the theory, but in the real-world implementations of such systems.
Such attacks are called Side-Channel attacks \cite{mai2011side, randolph2020power}.
Typical examples involve measuring power \cite{kocher1999differential}, electromagnetic emissions \cite{agrawal2003side}, optical emissions \cite{ferrigno2008aes, nassi2024video}, or execution time \cite{kocher1996timing}.

An example of a side-channel that affects QKD is that of an eavesdropper (henceforth denoted as Eve) having control over measurement devices, where Measurement-Device-Independent (MDI) QKD was proposed \cite{braunstein2012side, lo2012measurement} in response.
Eve could also externally interact with the QKD systems, for example, by injecting photons into the system to read information about the state encodings, known as a Trojan Horse attack \cite{lucamarini2015practical}, or by sending bright pulses of light to Avalanche Photodiode (APD) detectors in order to blind them, before then performing a faked-state attack \cite{koehler2018best}.
Another example of a side-channel attack on QKD systems is to measure the other degrees of freedom that are not being used to encode data \cite{nauerth2009information, biswas2021experimental}. 
Information from these measurements can then be used to estimate the prepared state.
Work has even been done into considering side-channel attacks against Continuous-Variable (CV) QKD \cite{derkach2016preventing, pereira2018hacking}.
For a more complete account see the review in \cite{xu2020secure} and the references therein.
\section*{Erroneous Raw Key Bit Ratio}

Prior to presenting the SD error recognition scheme, the type of error itself needs to be specified, as well as assumptions being made about the QKD scheme used to illustrate the results.

For the purposes of this paper it is assumed that the BB84 protocol is being used, where the Z and X bases are both used for key generation and parameter estimation, and are each used 50\% of the time.
Later on in this work it will be shown how other basis ratios besides 50:50 can be used by this scheme, relaxing this assumption.
It will also be assumed that there are four photodetectors that implement a passive detection scheme.
Figure 23 of \cite{gisin2002quantum} illustrates a passive QKD implementation.
Extension of this scheme to active QKD implementations is left for future work.

The type of error considered in this work is as follows: Bob will at some point in the execution of the QKD protocol be unable to receive one of the four states used by the BB84 protocol for the remainder of the QKD session.
This could occur in practice in a variety of ways, and these will be discussed in later sections. 

A SD error would result in one of the two bases only producing bits of one particular value, lowering the entropy of the raw key.
In the context of the assumed BB84 protocol, this would change the ratio of key bits from $2:2$, to $1:2$ or $2:1$ depending on which state is no longer being received by Bob. Later in this work it will be useful to think about the average bit value of the raw key, which for the above ratios are $1/2$, $2/3$, and $1/3$ respectively.
% \hl{As shown previously by equation \eqref{eq:intro_pa_compr_amount} and the following discussion, by lowering the entropy of the raw key, the size of the final secure key shared by Alice and Bob is reduced, which results in a lower overall key rate.}

To illustrate why the lowering of the entropy of the raw key is not desirable, consider the Devetak-Winter rate\cite{devetak2005distillation} for the SKR \mbox{$R=I(A:B)-I(A:E)$}.
Assuming an absolute best-case scenario, the term $I(A:E)$ can be ignored as it only detracts from the secret key rate. 
The first term $I(A:B)$ can be replaced by it's definition $I(A:B)=H(A)-H(A|B)$ and again, assuming the best-case scenario, the conditional entropy $H(A|B)$ becomes zero (ie. Alice and Bob's raw keys perfectly match).
As such, the entropy of Alice's raw key $H(A)$ can be taken to be an upper bound on Alice and Bob's mutual information, and by extension, the secret key rate:
\begin{equation}
    \label{eq:skr_dw_bound}
    R=I(A:B)-I(A:E)\leq I(A:B)\leq H(A)
\end{equation}
% \begin{equation}
%     \label{eq:dw_rate}
%     R=I(A:B)-I(A:E)
% \end{equation}
Therefore if the entropy of the raw key is reduced, the maximum possible secret key rate is also reduced.

\section*{Error Recognition}

In order for Alice and Bob to take action against a SD error, they must first be able to recognise it.
Alice could by means of classical communication tell Bob how many qubits were transmitted, and then Bob could calculate the loss using the number of qubits he received.
However this loss may have several causes, one of which could be a SD error, but it could also be due to an increase in the channel noise.
This approach doesn't yield sufficient information for Bob to make a determination as to the presence of a SD error.
As can be seen by the Binary Shannon Entropy:
\begin{equation}
    \label{eq:str_bin_shan_entr}
    H_2(p)=-p\log_2(p)-(1-p)\log_2(1-p)
\end{equation}
entropy is dependent on the probability distribution of the 0 and 1 bits in the raw key.
As a SD error decreases the entropy of the raw key, but general channel losses do not, estimating this probability distribution will allow Bob to make a determination as to the presence of the SD error.

If Bob were to sample the raw key bits live as they were obtained and calculated the mean of these values, he would see his calculated estimate of the mean deviate significantly from the nominal mean for the protocol.
For example, using the assumed BB84 protocol as stated above, the nominal mean is 0.5.
This problem is analogous to determining the bias of a biased coin, or more formally, determining the binomial parameter of a random Bernoulli variable.

However, in order to do this, two questions must now be answered:
\begin{enumerate}
    \item How many samples must Bob take in order to calculate a useful estimate of the mean?
    \item What does it mean for the estimated mean to deviate significantly?
\end{enumerate}
This first question is considered first, and by doing so will lead to a natural solution to the second question.

First, the actual mean of the distribution is defined as $\mu$, and Alice and Bob's estimate of the mean as:
\begin{equation}
    \label{eq:det_est_mean}
    \hat{\mu}=\frac{1}{n}\sum_{i=0}^nR_i
\end{equation}
where $n$ is the number of key bits sampled to calculate the estimate of the mean, and $R_i$ is the $i$-th bit sampled from the raw key.
The difference between the actual value and the estimated mean can then be quantified as $|\mu-\hat{\mu}|\leq\delta_\mu$ where $\delta_\mu>0$ is a parameter that specifies how close the estimate of the mean should be to the actual value, or rather how precise the estimate is.

However, as $\hat{\mu}$ is calculated using samples from a random variable, there is no guarantee that this scenario will hold.
As such, the following can be specified:
\begin{equation}
    \label{eq:det_scenario_eqn}
    P(|\mu-\hat{\mu}|\leq\delta_\mu)\leq1-\varepsilon,\quad0\leq\varepsilon\leq1
\end{equation}
where $\varepsilon$ is another parameter specifying that the scenario above holds with probability $1-\varepsilon$.

This scenario has been widely studied in the field of inferential statistics.
The Chernoff-Hoeffding bound provides a lower bound on the number of samples required \cite{chen2011exact}:
\begin{equation}
    \label{eq:det_chern_hoeff_bound}
    n\geq\frac{ln(2/\varepsilon)}{2{\delta_\mu}^2}.
\end{equation}

\begin{figure}[H]
    \centering
    \includegraphics[scale=0.75]{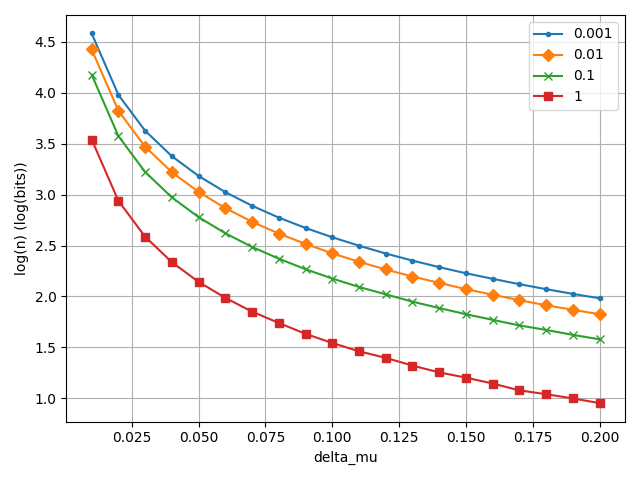}
    \caption{Logarithmic-scale plot of the number of bits from the Chernoff-Hoeffding bound against $\delta_\mu$ for different values of $\varepsilon$.}
    \label{fig:chern_hoeff_bound_contours}
\end{figure}

As can be seen from Fig.~\ref{fig:chern_hoeff_bound_contours}, the number of samples required is inversely proportional to both $\delta_\mu$ and $\varepsilon$, however $\delta_\mu$ scales particularly poorly.
% These $n$ samples used are applied to the latest $n$ bits received, and so this operation can be considered to be a sliding window over the raw key with a window width of $n$ bits.
These $n$ samples used are obtained from the latest $n$ bits received by Bob, and so this operation can be considered to be a sliding window over the raw key with a window width of $n$ bits.

\begin{figure}[H]
    \centering
    \includegraphics[scale=0.75]{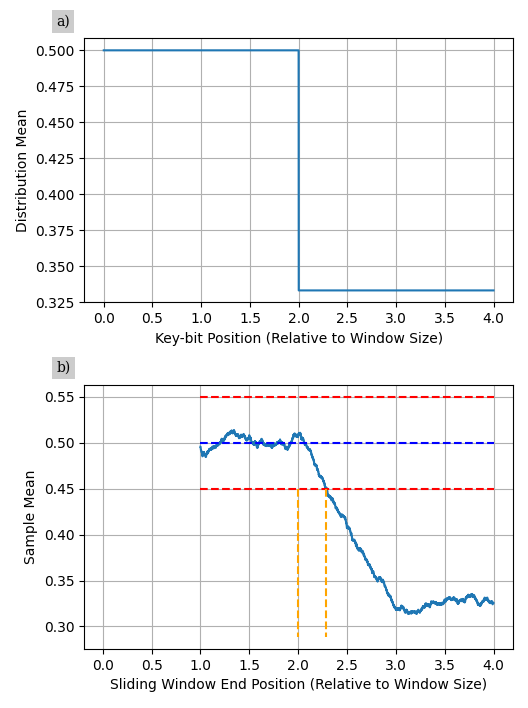}
    \caption{a.) the probability distribution used to create the random key. The distribution mean changes from $1/2$ to $1/3$ once the SD error has occurred. 
    b.) A simulation of Bob's estimated mean over time. The blue dashed line represents the nominal mean value, and the red dashed lines represent the upper and lower recognition thresholds. The first and second vertical orange dashed lines show where the SD error occurred, and where the recognition of the SD error was made respectively. This simulation was done with parameters $\delta_\mu=0.05$ and $\varepsilon=0.001$, hence a value for $t$ of $t=0.05$ was chosen as can be seen by the recognition thresholds.}
    \label{fig:det_reg_attack}
\end{figure}

Next is to consider the problem of finding some threshold after which it can be considered that the estimate of the mean has deviated significantly.
From here on this threshold shall be referred to as the recognition threshold and shall denoted by $t$.
The nominal mean shall be denoted by $\mu_N$.
By doing this, other values of the nominal mean are allowed besides the $1/2$ used in this work for illustrative purposes.
An example where a nominal mean besides $1/2$ arises is a variation of the BB84 protocol\cite{lo2005efficient} that does not use an equal proportion of bits from the Z and X bases, but rather uses some proportion $p$ of bits from the X basis such that $0<p\leq0.5$.
Then, by definition of $\delta_\mu$ it can be said that under nominal conditions 
% \begin{equation}
%     \label{eq:det_nom_est}
%     \mu_N-\delta_\mu\leq\hat{\mu}\leq \mu_N+\delta_\mu.
% \end{equation}
$\mu_N-\delta_\mu\leq\hat{\mu}\leq \mu_N+\delta_\mu$.
If the recognition threshold were to be set to less than $\delta_\mu$ then due to statistical fluctuations, there would be false positives outside of the specification provided by the parameters $\delta_\mu$ and $\varepsilon$, even under nominal conditions.
Therefore this imposes the bound of 
% \begin{equation}
%     \label{eq:det_thresh_bound}
%     t\geq\delta_\mu
% \end{equation}
$t\geq\delta_\mu$.
As such, the closest possible recognition threshold can be considered to be $t=\delta_\mu$.
A simulation a SD error and recognition of it \cite{young2024simulations} is shown in Fig.~\ref{fig:det_reg_attack}.

% 2.3385930309007232

The two questions that arose earlier have now been treated, but a new one has arisen in the meantime.
The derivations used to produce the Chernoff-Hoeffding bound make the Independent and Identically Distributed (IID) assumption.
Whilst Bob's samples are independent of one another, the probability distributions from which they are drawn are not identical in the presence of a SD error.

Under the IID assumption the presented sampling scheme can be described by equation~\eqref{eq:det_est_mean}, however when samples are taken from $k$ different binomial distributions then this can be rewritten as:
\begin{equation}
    \label{eq:det_non_iid_mean_est}
    \hat{\mu}=\frac{1}{n}\sum^{k-1}_{j=0}\sum^{n-1}_{i=0}x_{ji}
\end{equation}
where $x_{ji}$ is the $i$-th sample from the $j$-th probability distribution.

This difference means that the estimate will be as if it is from an identically distributed sample even though in the presence of the SD error it is not, and $\hat{\mu}$ will not truly reflect the mean of any one of the distributions, but rather present a weighted average based on the number of samples from each, given by
\begin{equation}
    \label{eq:mean_weight_avg}
    \hat{\mu}=\frac{1}{n}\sum_{i=0}^{k-1}\mu_in_i~.
\end{equation}

This means that the estimate of the mean will take time to move away from that of $\mu_N$ once the SD error has occurred.
This can be seen in Fig.~\ref{fig:det_reg_attack} by the sloped line once the probability distribution of raw key bits changes.
How far away from $\mu_N$ is tolerable can be bounded by 
% \begin{equation}
%     \label{eq:toler}
%     \mu_N-t\leq\hat{\mu}\leq \mu_N+t
% \end{equation}
$\mu_N-t\leq\hat{\mu}\leq \mu_N+t$.
This is illustrated in Fig.~\ref{fig:det_reg_attack} by the red dashed lines representing the upper and lower recognition thresholds.
\section*{Countermeasure}

Due to a SD error, there will be a region of the raw key that is partially insecure.
A two part countermeasure can be put in place.
The first part is to discard raw key bits that form the partially insecure region of the raw key.
The second part is to switch the protocol being used from the BB84 protocol to the 3-state BB84 protocol, which removes the dependence on the fourth state that is now no longer being received, and restores the raw key to having maximal entropy.

\subsection*{Discarding Compromised Key Bits}

In order to discard the required raw key bits, how many bits should be discarded needs to be quantified first.
First, the difference between the nominal mean $\mu_N$, and the current estimate of the mean $\hat{\mu}$ can be expressed.
Specifically the case where this difference first exceeds the recognition threshold $t$ needs to be considered, and is given by 
% \begin{equation}
%     \label{eq:res_diff_over_thresh}
%     |\mu_N-\hat{\mu}|=t~.
% \end{equation}
$|\mu_N-\hat{\mu}|=t$.
As the recognition thresholds are symmetric, and the rate of change in the estimated mean is the same when increasing or decreasing, only one case needs to be considered, in this case the decreasing case, where $\hat{\mu}<\mu_N$ such that 
% \begin{equation}
%     \label{eq:res_diff_over_thresh_no_abs}
%     \mu_N-\hat{\mu}=t~.
% \end{equation}
$\mu_N-\hat{\mu}=t$.
Then, by substituting in the definition for $\hat{\mu}$ given by equation~\eqref{eq:mean_weight_avg} and using just two probability distributions the following is obtained:
\begin{equation}
    \label{eq:res_det_pos_part_1}
    \mu_N-\frac{1}{n}(\mu_0n_0+\mu_1n_1)=t~.
\end{equation}
Here $\mu_N$, $\mu_0$, $\mu_1$, and $t$ are constants, and $n$, $n_0$, and $n_1$ are unknowns.
Here it is desirable to rewrite equation~\eqref{eq:res_det_pos_part_1} with $n_1$ as the subject to see how many bits it would take after the occurrence of the SD error for Bob to recognise the error.
The next step towards this is to normalise the number of samples such that $n=1$, giving $n_0+n_1=1$.
Substitutions can then be made into equation~\eqref{eq:res_det_pos_part_1} to eliminate $n$ and $n_0$, and can be rearranged to get:
\begin{equation}
    \label{eq:res_det_pos}
    n_1=\frac{\mu_N-t-\mu_0}{\mu_1-\mu_0}~.
\end{equation}
Explicitly defining two Bernoulli random variables $X_0=\{0.5, 0.5\}$ and $X_1=\{2/3, 1/3\}$ that represent the nominal case, and the case of the SD error respectively, as well as a nominal value of $\mu_N=0.5$ and a threshold $t=0.05$, a value of $n_1=0.3$ can be calculated.
This shows that on average, Bob will recognise a SD error after $0.3n$ bits into the insecure region, which looking at Fig.~\ref{fig:det_reg_attack} appears to be the case.

Hence, when Bob recognises a SD error, he could discard $0.3n$ bits worth of the raw key, however as this is only the case on average, sometimes he would not discard enough, and there would still be portions of insecure key remaining.
A naive solution to this would be to increase this value of $0.3n$ by some arbitrarily chosen multiplicative constant, large enough to discard enough of the raw key the vast majority of the time. 
However instead of an arbitrary constant a bit more of a sophisticated choice can be made.

\begin{figure}[H]
    \centering
    \includegraphics[scale=0.75]{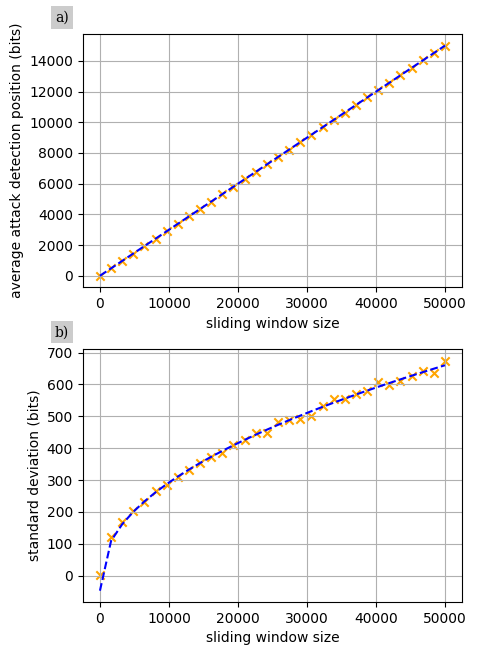}
    \caption{a.) This plot shows how the number of insecure raw key bits scales with the size of the sliding window.
    The dashed line shows the linear regression model that was fit to the dataset obtained via simulation.
    This coincides with the previously presented analytical result that showed that the proportion of insecure raw key bits scales linearly with the window size with a scaling factor of 0.3.
    b.) This plot shows the standard deviation of the results from the simulation.
    It was observed that this data exhibits a square-root scaling, and so a square-root regression model was fit to this data.
    The dashed line shows this regression model.
    For these simulations, the same parameters as with Fig.~\ref{fig:det_reg_attack} were used, and 2500 simulations were run for each window size.
    }
    \label{fig:res_disc_bits_plots}
\end{figure}

Simulations were run for different values of the sliding window size.
For each of these different values, many SD errors were simulated, and the time when each SD error was recognised was recorded.
The mean number of samples required for recognition ($n_r$), and the standard deviation were calculated.
Fig.~\ref{fig:res_disc_bits_plots} shows plots for these two quantities where linear and square root relationships were found for $n_r$ and the standard deviation respectively.

A linear regression was performed for the mean SD error recognition position in the sampling window, and a linear regression model of 
% \begin{equation}
%     \label{eq:res_det_pos_lin_reg}
%     n_r=\alpha n-\beta
% \end{equation}
$n_r=\alpha n-\beta$ was found, with $\alpha = 0.300$, $\beta = 5.058$, and a root mean squared error (RMSE) of 9.156 bits.
This model scales linearly with a scaling factor incredibly close to the value of $0.3$ as expected.

For the standard deviation it was noticed that it exhibited square root scaling behaviour, and so a square root regression model of 
% \begin{equation}
%     \label{eq:res_std_dev_reg}
%     \sigma=\sqrt{\alpha n-\beta}
% \end{equation}
$\sigma=\sqrt{\alpha n-\beta}$ was fit to this data, this time with $\alpha=8.770$, $\beta=2332.743$, and a RMSE of 11.068 bits.

From this, the size of the standard deviation can be calculated and multiples of it can be added onto the average error recognition position to get the number of bits that should be removed.

As an example, for a sampling window size of 50,000 bits, the average error recognition position is 14,993 bits, and the standard deviation is 661 bits \footnote{These quantities were rounded up to the next whole bit.}.
By using a confidence interval of three standard deviations, 99.73\% of the time enough bits will be discarded, and would result in 16976 bits needing to be discarded.

Now that the number of raw key bits to discard can be calculated, it is important to consider how discarding these affects the key rate.
If the length of the raw key is taken to be $l_r$ and the number of discarded raw key bits to be $n_d$, then the proportion of remaining raw key bits can be calculated, and the key rate prior to classical post-processing can be modified as such:
 \begin{equation}
     \label{eq:res_kr}
     R'=\frac{l_r-n_d}{l_r}R
 \end{equation}
\subsection*{Transitioning to 3-State BB84}

Once a SD error is recognised, it is known that one of the states can no longer be used.
As the BB84 protocol requires use of all four states, the naive approach would be to terminate the QKD session.
Whilst this would be a secure response to recognition of the SD error, this could cause downtime of potentially critical infrastructure.
It would be ideal if this could be avoided.

A more sophisticated approach would be to change the QKD protocol from BB84 to 3-State BB84, where the missing state in the 3-State BB84 specification corresponds to the state no longer being detected by Bob.
However Alice and Bob cannot just switch between protocols as their behaviour differs, albeit only slightly.

The first consideration is making Alice aware that she needs to switch to 3-State BB84, and which state needs to be removed.
On the face of it, revealing this information over the classical channel might seem like it presents a security concern.
However, in accordance with Kerckhoff's principle \cite{kerckhoffs2023cryptographie}, all parties including some eavesdropper Eve know the specifications of any ongoing protocol. 
Eve does not gain any more information than she would normally have if Alice and Bob were using 3-State BB84 from the outset.

The second consideration is what they need to change.
Alice needs to change her state generation procedure, and the instances used for parameter estimation to that of 3-State BB84.
It is worth noting that Alice does both of these changes, and Bob does not need to change anything about his part of the protocol. Alice and Bob can now continue to perform QKD and the uptime of the system has been maintained.
\section*{Motivating Examples}
\subsection*{Fault Detection}

Suppose in the passive detection scheme described earlier in this work that there was a complete failure of one of the four detectors.
Bob would no longer be able to detect one of the four BB84 states used by Alice.
This is a SD error.
The SD error recognition scheme presented above would be able to recognise this continued period of there being a fault in the system, and put in place the proposed countermeasure to maintain uptime of the QKD system.
This type of prolonged fault could also occur partially, for instance dust could build up in front of one of the detectors acting as an optical attenuator, therefore modifying the distribution mean of the key bits, resulting in a lowering of the entropy of the raw key.

This type of consideration is incredibly important for when QKD implementations move out of labs, and into real-world applications where experts are not immediately available to diagnose hardware faults.

\subsection*{Side-Channel Attack}

Suppose instead that some eavesdropper Eve had some side-channel attack capability that allows her to prevent Bob detecting one of the four BB84 states.
This could be achieved in a variety of ways, making this a very broad attack vector:
\begin{itemize}
    \item Physically obscure the input to, or disable one of the four detectors.
    \item Exploit a vulnerability in the software controlling the QKD system, eg. by telling Alice's setup to run the state-preparation according to the 3-State BB84 protocol, even though the user intends to use the BB84 protocol.
    \item Send relatively high power laser pulses into either Alice of Bob's setups in order to damage and/or permanently change the characteristics of their optical components, which can lead to vulnerabilities that can be exploited \cite{bugge2014laser, makarov2016creation}.
    \item Adjust the incident angle of the light on the input port to Bob's setup to change the path of the light such that it never reaches one of the detectors \cite{sajeed2015security}.
\end{itemize}

As discussed previously, by modifying the ratio of bit values in the raw key, Eve would be reducing the entropy of the raw key, leading to a reduction in the key rate.
As such, this side-channel attack is a form of denial-of-service (DoS) attack.

From the perspective of Bob this type of side-channel attack and hardware fault are indistinguishable.
Similar to how Bob cannot distinguish between channel noise and eavesdropping by Eve, and must therefore assume that the resulting Quantum Bit Error Rate is produced by an attack, Bob in this case must assume the worst case scenario: if a recognition of a SD error is made, then it is from a side-channel attack.
\section*{Conclusion}

This work began by defining a type of error, for which an error recognition scheme was devised using binomial parameter estimation.
Simulations were developed to show the operation of the recognition scheme, and certain characteristics of the scheme were investigated, such as how long it takes to recognise an error once it has begun.

A two-stage countermeasure was proposed that allows for continued operation of the QKD system after recognition of an error.
The first stage required discarding raw key bits that were generated after the error began, but before the error was recognised.
The second stage involves switching from BB84 to the 3-State BB84 protocol to remove the dependence on the state that is no longer being detected by Bob, allowing for a secure key to once again be generated.

Finally, there was discussion on how this error recognition scheme could be useful for applications such as fault detection of QKD implementation hardware, or a hypothetical type of side-channel attack.

Future work may look to further investigate different components within QKD implementations, how errors with those components result in changes within the results observed by Bob, and how autonomous recognition of these other types of errors could be done.
Future work could also consider that the Chernoff-Hoeffding bound used in this work is an upper bound, and not an exact solution which would allow for faster error recognition.
The analysis code produced for the simulations could also be applied to a real-world dataset gathered from experimental implementation to further validate the efficacy of the presented error recognition scheme.

\bibliography{refs}

\begin{thebibliography}{10}
\urlstyle{rm}
\expandafter\ifx\csname url\endcsname\relax
  \def\url#1{\texttt{#1}}\fi
\expandafter\ifx\csname urlprefix\endcsname\relax\def\urlprefix{URL }\fi
\expandafter\ifx\csname doiprefix\endcsname\relax\def\doiprefix{DOI: }\fi
\providecommand{\bibinfo}[2]{#2}
\providecommand{\eprint}[2][]{\url{#2}}

\bibitem{gisin2002quantum}
\bibinfo{author}{Gisin, N.}, \bibinfo{author}{Ribordy, G.}, \bibinfo{author}{Tittel, W.} \& \bibinfo{author}{Zbinden, H.}
\newblock \bibinfo{journal}{\bibinfo{title}{Quantum cryptography}}.
\newblock {\emph{\JournalTitle{Rev. Mod. Phys.}}} \textbf{\bibinfo{volume}{74}}, \bibinfo{pages}{145--195}, \doiprefix\url{10.1103/RevModPhys.74.145} (\bibinfo{year}{2002}).

\bibitem{pirandola2020advances}
\bibinfo{author}{Pirandola, S.} \emph{et~al.}
\newblock \bibinfo{journal}{\bibinfo{title}{Advances in quantum cryptography}}.
\newblock {\emph{\JournalTitle{Advances in optics and photonics}}} \textbf{\bibinfo{volume}{12}}, \bibinfo{pages}{1012--1236} (\bibinfo{year}{2020}).

\bibitem{shimizu2014performance}
\bibinfo{author}{Shimizu, K.} \emph{et~al.}
\newblock \bibinfo{journal}{\bibinfo{title}{Performance of long-distance quantum key distribution over 90-km optical links installed in a field environment of tokyo metropolitan area}}.
\newblock {\emph{\JournalTitle{Journal of Lightwave Technology}}} \textbf{\bibinfo{volume}{32}}, \bibinfo{pages}{141--151}, \doiprefix\url{10.1109/JLT.2013.2291391} (\bibinfo{year}{2014}).

\bibitem{amies-king2023feasibility}
\bibinfo{author}{Amies-King, B.} \emph{et~al.}
\newblock \bibinfo{title}{Feasibility of direct quantum communications between the uk and ireland via 224 km of underwater fibre}.
\newblock In \emph{\bibinfo{booktitle}{British and Irish Conference on Optics and Photonics 2023}}, \bibinfo{pages}{PS.13}, \doiprefix\url{10.1364/BICOP.2023.PS.13} (\bibinfo{publisher}{Optica Publishing Group}, \bibinfo{year}{2023}).

\bibitem{bennett2014quantum}
\bibinfo{author}{Bennett, C.~H.} \& \bibinfo{author}{Brassard, G.}
\newblock \bibinfo{journal}{\bibinfo{title}{Quantum cryptography: Public key distribution and coin tossing}}.
\newblock {\emph{\JournalTitle{Theoretical Computer Science}}} \textbf{\bibinfo{volume}{560}}, \bibinfo{pages}{7--11}, \doiprefix\url{https://doi.org/10.1016/j.tcs.2014.05.025} (\bibinfo{year}{2014}).
\newblock \bibinfo{note}{Theoretical Aspects of Quantum Cryptography – celebrating 30 years of BB84}.

\bibitem{fung2006security}
\bibinfo{author}{Fung, C.-H.~F.} \& \bibinfo{author}{Lo, H.-K.}
\newblock \bibinfo{journal}{\bibinfo{title}{Security proof of a three-state quantum-key-distribution protocol without rotational symmetry}}.
\newblock {\emph{\JournalTitle{Physical Review A}}} \textbf{\bibinfo{volume}{74}}, \bibinfo{pages}{042342} (\bibinfo{year}{2006}).

\bibitem{tamaki2014loss}
\bibinfo{author}{Tamaki, K.}, \bibinfo{author}{Curty, M.}, \bibinfo{author}{Kato, G.}, \bibinfo{author}{Lo, H.-K.} \& \bibinfo{author}{Azuma, K.}
\newblock \bibinfo{journal}{\bibinfo{title}{Loss-tolerant quantum cryptography with imperfect sources}}.
\newblock {\emph{\JournalTitle{Physical Review A}}} \textbf{\bibinfo{volume}{90}}, \bibinfo{pages}{052314} (\bibinfo{year}{2014}).

\bibitem{rusca2018security}
\bibinfo{author}{Rusca, D.}, \bibinfo{author}{Boaron, A.}, \bibinfo{author}{Curty, M.}, \bibinfo{author}{Martin, A.} \& \bibinfo{author}{Zbinden, H.}
\newblock \bibinfo{journal}{\bibinfo{title}{Security proof for a simplified bennett-brassard 1984 quantum-key-distribution protocol}}.
\newblock {\emph{\JournalTitle{Physical Review A}}} \textbf{\bibinfo{volume}{98}}, \bibinfo{pages}{052336} (\bibinfo{year}{2018}).

\bibitem{mai2011side}
\bibinfo{author}{Mai, K.}
\newblock \bibinfo{title}{Side channel attacks and countermeasures}.
\newblock In \emph{\bibinfo{booktitle}{Introduction to hardware security and trust}}, \bibinfo{pages}{175--194} (\bibinfo{publisher}{Springer}, \bibinfo{year}{2011}).

\bibitem{randolph2020power}
\bibinfo{author}{Randolph, M.} \& \bibinfo{author}{Diehl, W.}
\newblock \bibinfo{journal}{\bibinfo{title}{Power side-channel attack analysis: A review of 20 years of study for the layman}}.
\newblock {\emph{\JournalTitle{Cryptography}}} \textbf{\bibinfo{volume}{4}}, \bibinfo{pages}{15} (\bibinfo{year}{2020}).

\bibitem{kocher1999differential}
\bibinfo{author}{Kocher, P.}, \bibinfo{author}{Jaffe, J.} \& \bibinfo{author}{Jun, B.}
\newblock \bibinfo{title}{Differential power analysis}.
\newblock In \emph{\bibinfo{booktitle}{Advances in Cryptology—CRYPTO’99: 19th Annual International Cryptology Conference Santa Barbara, California, USA, August 15--19, 1999 Proceedings 19}}, \bibinfo{pages}{388--397} (\bibinfo{organization}{Springer}, \bibinfo{year}{1999}).

\bibitem{agrawal2003side}
\bibinfo{author}{Agrawal, D.}, \bibinfo{author}{Archambeault, B.}, \bibinfo{author}{Rao, J.~R.} \& \bibinfo{author}{Rohatgi, P.}
\newblock \bibinfo{title}{The em side—channel (s)}.
\newblock In \emph{\bibinfo{booktitle}{Cryptographic Hardware and Embedded Systems-CHES 2002: 4th International Workshop Redwood Shores, CA, USA, August 13--15, 2002 Revised Papers 4}}, \bibinfo{pages}{29--45} (\bibinfo{organization}{Springer}, \bibinfo{year}{2003}).

\bibitem{ferrigno2008aes}
\bibinfo{author}{Ferrigno, J.} \& \bibinfo{author}{Hlav{\'a}{\v{c}}, M.}
\newblock \bibinfo{journal}{\bibinfo{title}{When aes blinks: introducing optical side channel}}.
\newblock {\emph{\JournalTitle{IET Information Security}}} \textbf{\bibinfo{volume}{2}}, \bibinfo{pages}{94--98} (\bibinfo{year}{2008}).

\bibitem{nassi2024video}
\bibinfo{author}{Nassi, B.} \emph{et~al.}
\newblock \bibinfo{title}{Video-based cryptanalysis: Extracting cryptographic keys from video footage of a device’s power led captured by standard video cameras}.
\newblock In \emph{\bibinfo{booktitle}{2024 IEEE Symposium on Security and Privacy (SP)}}, \bibinfo{pages}{163--163} (\bibinfo{organization}{IEEE Computer Society}, \bibinfo{year}{2024}).

\bibitem{kocher1996timing}
\bibinfo{author}{Kocher, P.~C.}
\newblock \bibinfo{title}{Timing attacks on implementations of diffie-hellman, rsa, dss, and other systems}.
\newblock In \emph{\bibinfo{booktitle}{Advances in Cryptology—CRYPTO’96: 16th Annual International Cryptology Conference Santa Barbara, California, USA August 18--22, 1996 Proceedings 16}}, \bibinfo{pages}{104--113} (\bibinfo{organization}{Springer}, \bibinfo{year}{1996}).

\bibitem{braunstein2012side}
\bibinfo{author}{Braunstein, S.~L.} \& \bibinfo{author}{Pirandola, S.}
\newblock \bibinfo{journal}{\bibinfo{title}{Side-channel-free quantum key distribution}}.
\newblock {\emph{\JournalTitle{Physical review letters}}} \textbf{\bibinfo{volume}{108}}, \bibinfo{pages}{130502} (\bibinfo{year}{2012}).

\bibitem{lo2012measurement}
\bibinfo{author}{Lo, H.-K.}, \bibinfo{author}{Curty, M.} \& \bibinfo{author}{Qi, B.}
\newblock \bibinfo{journal}{\bibinfo{title}{Measurement-device-independent quantum key distribution}}.
\newblock {\emph{\JournalTitle{Physical review letters}}} \textbf{\bibinfo{volume}{108}}, \bibinfo{pages}{130503} (\bibinfo{year}{2012}).

\bibitem{lucamarini2015practical}
\bibinfo{author}{Lucamarini, M.} \emph{et~al.}
\newblock \bibinfo{journal}{\bibinfo{title}{Practical security bounds against the trojan-horse attack in quantum key distribution}}.
\newblock {\emph{\JournalTitle{Physical Review X}}} \textbf{\bibinfo{volume}{5}}, \bibinfo{pages}{031030} (\bibinfo{year}{2015}).

\bibitem{koehler2018best}
\bibinfo{author}{Koehler-Sidki, A.} \emph{et~al.}
\newblock \bibinfo{journal}{\bibinfo{title}{Best-practice criteria for practical security of self-differencing avalanche photodiode detectors in quantum key distribution}}.
\newblock {\emph{\JournalTitle{Physical Review Applied}}} \textbf{\bibinfo{volume}{9}}, \bibinfo{pages}{044027} (\bibinfo{year}{2018}).

\bibitem{nauerth2009information}
\bibinfo{author}{Nauerth, S.}, \bibinfo{author}{F{\"u}rst, M.}, \bibinfo{author}{Schmitt-Manderbach, T.}, \bibinfo{author}{Weier, H.} \& \bibinfo{author}{Weinfurter, H.}
\newblock \bibinfo{journal}{\bibinfo{title}{Information leakage via side channels in freespace bb84 quantum cryptography}}.
\newblock {\emph{\JournalTitle{New Journal of Physics}}} \textbf{\bibinfo{volume}{11}}, \bibinfo{pages}{065001} (\bibinfo{year}{2009}).

\bibitem{biswas2021experimental}
\bibinfo{author}{Biswas, A.}, \bibinfo{author}{Banerji, A.}, \bibinfo{author}{Chandravanshi, P.}, \bibinfo{author}{Kumar, R.} \& \bibinfo{author}{Singh, R.~P.}
\newblock \bibinfo{journal}{\bibinfo{title}{Experimental side channel analysis of bb84 qkd source}}.
\newblock {\emph{\JournalTitle{IEEE Journal of Quantum Electronics}}} \textbf{\bibinfo{volume}{57}}, \bibinfo{pages}{1--7} (\bibinfo{year}{2021}).

\bibitem{derkach2016preventing}
\bibinfo{author}{Derkach, I.}, \bibinfo{author}{Usenko, V.~C.} \& \bibinfo{author}{Filip, R.}
\newblock \bibinfo{journal}{\bibinfo{title}{Preventing side-channel effects in continuous-variable quantum key distribution}}.
\newblock {\emph{\JournalTitle{Physical Review A}}} \textbf{\bibinfo{volume}{93}}, \bibinfo{pages}{032309} (\bibinfo{year}{2016}).

\bibitem{pereira2018hacking}
\bibinfo{author}{Pereira, J.} \& \bibinfo{author}{Pirandola, S.}
\newblock \bibinfo{journal}{\bibinfo{title}{Hacking alice's box in continuous-variable quantum key distribution}}.
\newblock {\emph{\JournalTitle{Physical Review A}}} \textbf{\bibinfo{volume}{98}}, \bibinfo{pages}{062319} (\bibinfo{year}{2018}).

\bibitem{xu2020secure}
\bibinfo{author}{Xu, F.}, \bibinfo{author}{Ma, X.}, \bibinfo{author}{Zhang, Q.}, \bibinfo{author}{Lo, H.-K.} \& \bibinfo{author}{Pan, J.-W.}
\newblock \bibinfo{journal}{\bibinfo{title}{Secure quantum key distribution with realistic devices}}.
\newblock {\emph{\JournalTitle{Reviews of modern physics}}} \textbf{\bibinfo{volume}{92}}, \bibinfo{pages}{025002} (\bibinfo{year}{2020}).

\bibitem{devetak2005distillation}
\bibinfo{author}{Devetak, I.} \& \bibinfo{author}{Winter, A.}
\newblock \bibinfo{journal}{\bibinfo{title}{Distillation of secret key and entanglement from quantum states}}.
\newblock {\emph{\JournalTitle{Proceedings of the Royal Society A: Mathematical, Physical and engineering sciences}}} \textbf{\bibinfo{volume}{461}}, \bibinfo{pages}{207--235} (\bibinfo{year}{2005}).

\bibitem{chen2011exact}
\bibinfo{author}{Chen, X.}
\newblock \bibinfo{journal}{\bibinfo{title}{Exact computation of minimum sample size for estimation of binomial parameters}}.
\newblock {\emph{\JournalTitle{Journal of Statistical Planning and Inference}}} \textbf{\bibinfo{volume}{141}}, \bibinfo{pages}{2622--2632} (\bibinfo{year}{2011}).

\bibitem{lo2005efficient}
\bibinfo{author}{Lo, H.-K.}, \bibinfo{author}{Chau, H.~F.} \& \bibinfo{author}{Ardehali, M.}
\newblock \bibinfo{journal}{\bibinfo{title}{Efficient quantum key distribution scheme and a proof of its unconditional security}}.
\newblock {\emph{\JournalTitle{Journal of Cryptology}}} \textbf{\bibinfo{volume}{18}}, \bibinfo{pages}{133--165} (\bibinfo{year}{2005}).

\bibitem{young2024simulations}
\bibinfo{author}{Young, M.}, \bibinfo{author}{Lucamarini, M.} \& \bibinfo{author}{Pirandola, S.}
\newblock \bibinfo{title}{State-blocking side-channel attack simulations}, \doiprefix\url{10.5281/zenodo.13738092} (\bibinfo{year}{2024}).

\bibitem{kerckhoffs2023cryptographie}
\bibinfo{author}{Kerckhoffs, A.}
\newblock \emph{\bibinfo{title}{La cryptographie militaire}} (\bibinfo{publisher}{BoD--Books on Demand}, \bibinfo{year}{2023}).

\bibitem{bugge2014laser}
\bibinfo{author}{Bugge, A.~N.} \emph{et~al.}
\newblock \bibinfo{journal}{\bibinfo{title}{Laser damage helps the eavesdropper in quantum cryptography}}.
\newblock {\emph{\JournalTitle{Phys. Rev. Lett.}}} \textbf{\bibinfo{volume}{112}}, \bibinfo{pages}{070503}, \doiprefix\url{10.1103/PhysRevLett.112.070503} (\bibinfo{year}{2014}).

\bibitem{makarov2016creation}
\bibinfo{author}{Makarov, V.} \emph{et~al.}
\newblock \bibinfo{journal}{\bibinfo{title}{Creation of backdoors in quantum communications via laser damage}}.
\newblock {\emph{\JournalTitle{Phys. Rev. A}}} \textbf{\bibinfo{volume}{94}}, \bibinfo{pages}{030302}, \doiprefix\url{10.1103/PhysRevA.94.030302} (\bibinfo{year}{2016}).

\bibitem{sajeed2015security}
\bibinfo{author}{Sajeed, S.} \emph{et~al.}
\newblock \bibinfo{journal}{\bibinfo{title}{Security loophole in free-space quantum key distribution due to spatial-mode detector-efficiency mismatch}}.
\newblock {\emph{\JournalTitle{Physical Review A}}} \textbf{\bibinfo{volume}{91}}, \bibinfo{pages}{062301} (\bibinfo{year}{2015}).

\end{thebibliography}

\section*{Acknowledgements}
S.P. acknowledges support from the EPSRC via the UK Quantum Communications Hub (Grant No. EP/T001011/1).

\section*{Author contributions}
M.Y conceived of and conducted the research, and prepared the manuscript. S.P and M.L supervised the project. All authors reviewed and contributed to the revision of the manuscript.

\section*{Declarations}

\section*{Competing Interests}
The authors declare no competing interests.

\section*{Additional information}

\end{document}